\documentclass[aps,showpacs,nobibnotes,amsmath,amssymb,prd]{revtex4}

\begin{document}
\draft{}
\preprint{}
\title{ A Four Dimensional  Superstring from the Bosonic
      String with Some Applications.}
\author{B. B. Deo and L. Maharana\footnote{E-Mail:lmaharana@iopb.res.in}}
\affiliation{Physics Department, Utkal University,Bhubaneswar-751004, India.}
\begin{abstract}
A string in four dimensions is constructed by supplementing it with  
forty four Majorana fermions. The later are represented by eleven
vectors in the bosonic representation $SO(D-1,1)$. The central charge is 26.
The fermions are grouped in
such a way that the resulting action is world sheet supersymmetric.
The energy momentum and current generators satisfy the super-Virasoro
algebra.  GSO projections are necessary
for proving modular invariance. Space-time supersymmetry algebra is
deduced and is substantiated for specific modes of zero mass. The
symmetry group of the model can descend to the low energy standard
model group $SU (3) \times SU_L (2) \times U_Y (1)$ through 
the Pati-Salam group.
\end{abstract}
\maketitle

\section{Introduction}
String theory was invented \cite{bd1} as a sequel to dual
resonance models \cite{bd2} to explain the properties of 
strongly interacting particles in four dimensions. Assuming
a background gravitational field and demanding Weyl invariance, 
the Einstein equations of general relativity could be deduced. 
It was believed that about these classical solutions, one can 
expand and find the quantum corrections. But difficulties arose 
at the quantum level. Eventhough the strong interaction amplitude 
obeyed crossing, it was no longer unitary. There were anomalies 
and ghosts. Due to these compelling reasons it was necessary for 
the open string to live in 26 dimensions \cite{bd3,bd10}. At 
present the most successful theory is a ten dimensional superstring 
on a Calabi-Yau manifold or an orbifold. However, in order to 
realise the programme of the string unification of all the four 
types of interactions, one must eventually arrive at a
theory in four flat space-time dimensions, with N=1 supersymmetry
and chiral matter fields. Recently one of us~\cite{30} has proposed
a new type of four dimensional Superstring with those
features. Certain questions like equivalence of light cone action to
the Superconformal ghost action were raised. This paper is an attempt 
in elaborating this letter. Considerable research, relating to this 
paper had also been done by Gates and his collaborators~\cite{gates}.

A lot of research has been done to construct four 
dimensional strings  \cite{bd99}, specially in the 
latter half of the eighties. Antoniadis
et al\cite{ia} have constructed a four dimensional superstring supplemented 
by eighteen real fermions in trilinear coupling. The central charge of the
construction is 15. Chang and Kumar\cite{ia} have discussed the
problem with Thirring fermions. Kawai et al \cite{ia} have also considered 
        four dimensional models in a different context
than the model proposed here. None of these models makes direct
contact with the standard  model.  There are no superconformal ghosts
as well. Their contribution of eleven to the central charge is made up 
by those coming from the Lorentzian fermionic ghosts. This will be
shown in section III using light cone gauge, with detailsin section VIII.

Consistent superstrings as solutions of $D=26$ bosonic string has been 
shown to exist by Casper, Engler, Nicolai and Taormina \cite{6}. Quite
recently Englert, Hourat and Taormina \cite{7} have extended this work
to brane fusion in the bosonic string resulting in a fermionic
string. The present work, which has similar purpose, is quite different 
and novel. We search for supersymmetry within the compactified
Nambu-Goto string in 26 dimensions by introducing eleven four-plet of 
Majorana fermions transforming in the vector representation of the
Lorentz Group $SO(3,1)$ and arrive at a four, not a ten dimensional
superstring\cite{30}.

The string, we construct, is essentially, the 26 dimensional ordinary bosonic
string in which the bosonic coordinates $X^{\mu}$ are restricted to be four
with $\mu = 0, 1, 2, 3$.  We show
that an action can be constructed which is 
supersymmetric. All the relevant algebra given in standard text books
follow. But subtle differences, which have to be pointed out, are 
repeated in the paper.

Using vectorial fermions for compactification introduces eleven
longitudinal ghost modes. Fortunately there are exactly eleven
component subsidiary conditions to eliminate them and make the theory
physical. Complication due to so many vectors makes a light cone
analysis impracticable. We have to fall back on the covariant
formation of superstring theory. The central charge is calculatred to
be 26.

In section II, we give the details of the world sheet supersymmetric
model. The following section III, the local $2D$ and local $4D$ supersymmetric 
actions are discussed. Section IV gives the usual quantization and 
super-Virasoro algebra isdeduced in the section V. Bosonic states are 
constructed in Section VI. Fadeev- Popov ghosts are introduced and the
BRST charge is explicitly given in section VII and VIII. Ramond states 
have been worked out in section IX. In section X,   
the mass spectrum of the model and the necessary GSO projections to
eliminate the half integral spin states are introduced.  In section
XI, we show that these projections are necessary to prove the modular 
invariance of the model. Space-time supersymmetry algebra is
satisfied and is shown to exist for the generated zero mass modes in
section XII. In section XIII we show how the  
chain $SO(44)\to SO(11)\to SO(6)\times SO(5) \to SU(4) \times SU(2)
\times SU(2)$ from the string theory and then descend to $SU_C (3)\times SU_L(2)
\times U_Y(1)$. We calculate that the stringy Pati-Salam group 
$SU(4) \times SU_L (2) \times SU_R (2)$ 
breaks at an intermediate mass $M_R \simeq 5 \times 10^{14}$ GeV
giving the left-handed 
neutrino a small mass, which has now been observed in the top
sector. An action is written down with a curved metric in Section XIV
and is world sheet supersymmetric. The vanishing of the one loop
$\beta$-function results in  the vanishing of the Ricci tensor in the
normal one time, three space universe.

The literature on string theory is very vast and exist in most
text books on the subject. The references serve only as a guide
to elucidate the model. 

\section{The Model} 
As stated earlier, the model consists of  replacing the 26
vector bosons of the 26-d bosonic string by the 4 bosonic cordinates
of the four dimensions and 44  Majorana fermions equivalent to the
remaining 22 bosonic co-ordinates \cite{bd11}. Traditionally, these
compactifying fermions are world sheet scalars. The novel feature of
this model is that we shall take them as Lorentz vectors in  the
bosonic representation, $SO(D-1,1)$.

We divide the fermions into four 
groups. They are labelled by the space time indices $\mu=0,1,2,3$. 
Each group contains 11 fermions.  We begin to search for a four vector 
which is also an anticommutating Majorana field $\Psi^\mu$ to
superpartner $X^\mu$. As a first step, the 11 fermions are 
divided into two groups, one containing six and the other five
fermions.  We level one group of real Majorana fields by $\psi^{\mu,j}$ with
$j=1,2,3,4,5,6$ and the other  by $\phi^{\mu, k}_R$ with $k=1,2,3,4,5 =
j-6$. To differentiate and for convinience, we use $\phi^{\mu, k} = i
\phi_R^{\mu,k}$ and scrupulously keep track of the factor `i'
everywhere; there is an arbitary phase factor in defining a congugate
to the Majorana spinor. At this stage, we introduce  vector like
objects  ${\bf e}_{\psi}$
and ${\bf e}_{\phi}$ with eleven components, e.g. for $j = 3$ and $k =
3$ with $\psi$ and $\phi$ as book keeping suffixes,

\begin{equation}
{\bf e}_{\psi}^3 = ( 0, 0, 1, 0, 0, 0, 0, 0; 0, 0, 0 )
\end{equation}
and
\begin{equation}
{\bf e}_{\phi}^3 = ( 0, 0, 0, 0, 0, 0; 0, 0, 1, 0, 0 )
\end{equation}
with the properties $e_{\psi}^{j} e_{\psi, j^\prime} =
\delta^{j}_{j^\prime}$, $e_{\phi}^{k} e_{\phi , k^\prime} =
\delta^{k}_{k^\prime}$ and ${\bf e}_{\phi} \cdot {\bf e}_{\psi} = 0$.
We shall frequently use the relation ${e}_\psi^j {e}_{\psi,j}
= 6$ and ${e}_\phi^k e_{\phi, k} = 5$.The upper
index refers to a column, and the lower to a row.

The string action of the model is 
\begin{equation}
S=-\frac{1}{2 \pi} \int d^2 \sigma \left[ {\partial}_{\alpha} 
X^{\mu}{\partial}^{\alpha} X_{\mu}-i \bar{\psi}^{\mu,j}
\rho^{\alpha} {\partial}_{\alpha}\psi_{\mu,j}-i \bar{\phi}^{\mu,k} 
\rho^{\alpha}{\partial}_{\alpha}\phi_{\mu,k}\right]\;,
\end{equation}

$\rho^{\alpha}$ are the two dimensional Dirac matrices
\begin{eqnarray}
\rho^0 =
\left ( 
\begin{array}{cc}
0~~~~~~ -i\\
i~~~~~~~  0\\
\end{array}
\right )
\end{eqnarray}
\begin{eqnarray}
\rho^1 =
\left (
\begin{array}{cc}
0~~~~~~ i\\
i~~~~~~  0\\
\end{array}
\right ),
\end{eqnarray}
and obey 
\begin{equation}
\{\rho^{\alpha}, \rho^{\beta}\}=-2 \eta^{\alpha \beta}.
\end{equation}

Further 
\begin{equation}
\bar{\psi} = \tilde{\psi} \rho^0 ;~~~~~~~~~~ \bar{\phi} = \tilde{\phi} \rho^0.
\end{equation}
Such an action had also been written down by Gates et al~\cite{gates}.
 
In general we follow the notations and conventions of reference 
\cite{bd5} whenever omitted by us. $X^{\mu}(\sigma,\tau)$ are 
the string coordinates. The  fermions $\psi'$s and
$\phi$'s are decomposed as
\begin{equation}
\psi=\left( \begin{array}{c} \psi_-\\ \psi_+ \end{array}
\right)\;,
\;\;\;\mbox{and}\;\;\;\; 
\phi=\left( \begin{array}{c}\phi_-\\ \phi_+ \end{array}
\right)\;.
\end{equation}

The action is found to be invariant under the following infinitesimal
transformations 
\begin{equation}
\delta X^{\mu}=\bar{\epsilon} \; \left (  e_\psi^j
  \psi^{\mu}_{j} +i  e_\phi^k \phi^{\mu}_{k} \right ) = (e_\psi^j
\bar{\psi}_{j}^{\mu} + i e_\phi^k \bar{\phi}_k^\mu ) \epsilon,
\end{equation}
\begin{equation}
\delta\psi^{\mu,\;j}=-i e_\psi^j \rho^{\alpha}\; \partial_{\alpha} X^{\mu}
\;\epsilon, ~~~~~~~~ \delta \bar{\psi}^{\mu, j} = i e_{\psi}^j
\bar{\epsilon} \rho^\alpha \partial_\alpha X^\mu,
\end{equation}
and
\begin{equation}
\delta \phi^{\mu,\;k}= e_\phi^k \rho^{\alpha}\;\partial_{\alpha}\;
X^{\mu} \;\epsilon, ~~~~~~~~~ \delta \bar{\phi}^{\mu, k} = - e_\phi^{k}
\bar{\epsilon} \rho^{\alpha} \partial_\alpha X^\mu.
\end{equation}
 $\epsilon$ is an infinitesimally small constant anticommuting 
Majorana spinor. The commutator of the two supersymmetry 
transformations gives a spatial translation, namely 
\begin{equation}
[\delta_1,\delta_2] X^{\mu}=a^{\alpha}\;{\partial}_{\alpha}X^{\mu},
\end{equation}
\begin{equation}
[\delta_1,\delta_2] \psi^{\mu, j}=a^{\alpha} {\partial}_{\alpha}
\psi^{\mu, j},
\end{equation}
and
\begin{equation}
[\delta_1 , \delta_2] \phi^{\mu, k} = a^\alpha \partial_\alpha
\phi^{\mu, k},
\end{equation}
where 
\begin{equation}
a^{\alpha}=2i\;\bar \epsilon_1\; \rho^{\alpha}\; \epsilon_2.
\end{equation}
In deriving these, the Dirac equations for the spinors have been
used. It is interesting to note that  we must also choose 
$\psi_j^\mu = e_{\psi, j} \Psi^\mu$ and $\phi_k^\mu = i e_{\phi,k} 
\Psi^\mu$ leading to the desired 
\begin{equation}
\Psi^{\mu} = 6\Psi^\mu - 5\Psi^\mu = e_{\psi}^{j} \psi^{\mu}_{j}+i
e_{\phi}^k \phi^{\mu}_{k} .\label{a1}
\end{equation}
We can recast the infinitesmal transformations as 
\begin{equation}
\delta X^\mu = \bar{\epsilon} \Psi^\mu,\label{a2}
\end{equation}
\begin{equation}
\delta \Psi^\mu = - i \rho^\alpha \partial_\alpha X^\mu \epsilon\label{a3}
\end{equation}
and
\begin{equation}
[ \delta_1 , \delta_2 ] \Psi^\mu = a^\alpha \partial_\alpha \Psi^\mu.\label{a4}
\end{equation}

The nonvanishing equal time commutator and anticommutators are
\begin{equation}
\left [\partial_{\pm} X^\mu (\sigma, \tau), \partial_{\pm} X^\nu
  (\sigma^\prime, \tau) \right ] = \pm \frac{\pi}{2} \eta^{\mu\nu}
\delta^\prime(\sigma-\sigma^\prime)
\end{equation}
and
\begin{equation}
\{ \psi^\mu_A (\sigma, \tau), \psi_B^\nu (\sigma^\prime , \tau) \} =
\pi \delta (\sigma - \sigma^\prime) \eta^{\mu \nu} \delta_{AB}.
\end{equation}
Even though
\begin{equation}
\{ \phi_A^\mu (\sigma, \tau), \phi_B^\nu (\tau) \} = \pi \delta
(\sigma - \sigma^\prime) \eta^{\mu \nu} \delta_{AB},
\end{equation}
there are no ghost quanta other than $\mu = \nu = 0$ due to a negative
creation operator phase. 

Also we have,
\begin{equation}
\{ \Psi_A^\mu (\sigma, \tau), \Psi_B^\nu (\sigma^\prime, \tau) \} =
\pi \delta (\sigma - \sigma^\prime) \eta^{\mu \nu} \delta_{AB}.
\end{equation}
The Noether super-current is

\begin{equation}
J_{\alpha}=\frac{1}{2} \rho^{\beta}\;\rho_{\alpha}
\;\Psi^{\mu}\; \partial_{\beta} X_{\mu}.
\end{equation}

We now follow the standard procedure. The lightcone components 
of the current and energy momentum tensors are
\begin{equation}
J_+=\partial_+X_{\mu}\; \Psi^{\mu}_+ ,
\end{equation}
\begin{equation}
J_-=\partial_-X_{\mu}\;\Psi^{\mu}_- ,
\end{equation}
\begin{equation}
T_{++}=\partial_+X^{\mu} \partial_+X_{\mu}+\frac{i}{2}
\psi_+^{\mu,\;j}\;\partial_+\psi_{+\mu,\;j}+\frac{i}{2}
\phi_+^{\mu,\;k}\;\partial_+\phi_{+\mu,\;k},\label{e}
\end{equation}
and
\begin{equation}
T_{--}=\partial_-X^{\mu} \partial_-X_{\mu}+\frac{i}{2}
\psi_-^{\mu,\;j}\partial_-\psi_{-\mu,\;j}+\frac{i}{2}
\phi_-^{\mu,\;k}\partial_-\phi_{-\mu,\;k}.\label{f}
\end{equation} 
where $\partial_{\pm}=\frac {1}{2} (\partial_{\tau}\pm 
\partial_{\sigma})$.

Next we calculate the following two commutators at equal $\tau$,


\begin{equation}
\left [ T_{+ +} (\sigma), T_{+ +} (\sigma^\prime) \right ]
= i \pi \delta^\prime (\sigma - \sigma^\prime )~~ \left ( T_{+ +}
(\sigma) + T_{+ +} (\sigma^{\prime}) \right ),
\end{equation}
and
\begin{equation}
\left [ T_{+ +} (\sigma), J_{+} (\sigma^\prime) \right ]
= i \pi \delta^\prime (\sigma - \sigma^\prime ) \left ( J_{+}
(\sigma) + \frac{1}{2} J_{+} (\sigma^{\prime}) \right ).
\end{equation}

To satisty the Jacobi Identity
\begin{equation}
\left [ T_{++} (\sigma),\left\{ J_+ (\sigma^\prime), J_+ (\sigma^{\prime
  \prime}) \right\}\right ] = \{ [ T_{++} (\sigma), J_{+}(\sigma^{\prime})],
J_+(\sigma^{\prime \prime})\} + (\sigma^\prime \leftrightarrow
\sigma^{\prime \prime} ).
\end{equation}
We use ~~~~$\delta(\sigma - \sigma^\prime) \delta^\prime
(\sigma-\sigma^{\prime \prime}) + \delta^\prime (\sigma-\sigma^\prime) 
\delta (\sigma-\sigma^{\prime \prime}) = \delta(\sigma^\prime -
\sigma^{\prime \prime}) \delta^\prime (\sigma - \sigma^\prime)$
~~~~~and verify that
\begin{eqnarray}
\{ J_{+} (\sigma), J_{+}(\sigma^{\prime})\} && = \pi
\delta(\sigma - \sigma^{\prime}) T_{++} (\sigma).  \nonumber \\
{\rm Similarly}~~~~~~~~~~~~~~ \{ J_{-} (\sigma),
J_{-}(\sigma^{\prime})\} 
&& = \pi\delta(\sigma - \sigma^{\prime}) T_{--} (\sigma) \nonumber \\
{\rm and}~~~~~~~~~~~~~ \{ J_{+} (\sigma), J_{-}(\sigma^{\prime})\} 
&& = 0 .\label{g}
\end{eqnarray}
Interestingly the sum of the two groups of fermionic terms in 
the equivalent Nambu-Goto action (3) becomes
\[ \psi^{\mu, j} \rho^\alpha \partial_\alpha \psi_{\mu, j} +
\phi^{\mu, k} \rho^\alpha \partial_\alpha \phi_{\mu, k} = 6
\bar{\Psi}^\mu \rho^\alpha \partial_\alpha \Psi_\mu - 5 \bar{\Psi}^\mu 
\rho^\alpha \partial_\alpha \Psi_\mu = \bar{\Psi}^\mu \rho^\alpha
\partial_\alpha \Psi_\mu . \]
We obtain the reduced four dimensional superstring action
\begin{equation}
S = - \frac{1}{2 \pi} \int d^2 \delta ( \partial^\alpha X^\mu
\partial_\alpha X_\mu - i \bar{\Psi}^\mu \rho^\alpha \partial_\alpha
\Psi_\mu ).
\end{equation}
Similarly, the energy momentum tensor equations (\ref{e}) and
(\ref{f}) 
can be cast in the form containing $\Psi$'s only and the current
anticommutators are deduced easily and directly without using the
Jacobi Identity. The algebra closes.

In view of equation  (\ref{g}), we  may just postulate that\cite{bd5}
\begin{equation}
 0 = J_{+} = J_{-} = T_{++} = T_{--} .\label{a5}
\end{equation}
We also  obtain the component constraints
\[ \partial_\pm X_\mu \psi_\pm^{\mu, j} = \partial_\pm X_\mu e_\psi^j
\Psi_\pm^\mu = 0 , \; \; j = 1, 2, \cdots 6 
\]
and
\begin{equation}
\partial_\pm X_\mu \phi_{\pm}^{\mu, k} = \partial_\pm X_\mu
e_\phi^k \Psi^\mu_\pm = 0 , \; \; k = 1, 2, \cdots 5
\end{equation}
These subsidiary postulates enable us to construct null physical states
eliminating all negative norm states. It also follows that 
for the two groups of fermions, the two Lorentz and group invariant 
constraint equations are
\begin{equation}
\delta_\pm X_\mu  e_\psi^j \psi_{\pm j}^{\mu} =0 \label{b1}
\end{equation}
and
\begin{equation}
\delta_\pm X_\mu  e_\psi^k \phi_{\pm k}^{\mu} = 0\label{b2}
\end{equation}

\section{Local Two Dimensional and Four Dimensional supersymmetry}
In view of equations (\ref{a1}) to (\ref{a2}), $(X^\mu , \Psi^\mu)$ behave like a
supersymmetric pair. In addition to this pair, we introduce a
`zweibein'  $e_\alpha (\sigma,\tau)$ and its supersymmetric partnet
$\chi_\alpha$ into the theory. This $\chi_\alpha$ is a two dimensional 
spinor as well as a two dimensional vector. Following reference [9],
we deduce the local 2D supersymmetric action.
\begin{eqnarray}
S = && - \frac{1}{2\pi} \int d^2 \sigma~e~ \{ h^{\alpha \beta}
\partial_\alpha X^\mu \partial_\beta X^\mu - i(\bar{\psi}^{\mu,j}
\rho^{\alpha} \partial_\alpha \psi_{\mu,j} + \bar{\phi}^{\mu,k}
\rho^\alpha \partial_\alpha \phi_{\mu, k} ) \nonumber \\
&& + 2 \bar{\chi}_\alpha \rho^\beta \rho^\alpha \Psi^\mu
\partial_\beta X^\mu + \frac{1}{2} \bar{\Psi}^\mu \Psi_\mu
\bar{\chi}_\alpha \rho^\beta \rho^\alpha \chi_\alpha \}
\end{eqnarray}
which is invariant under local supersymmetric transformations
\[ \delta X^\mu = \bar{\epsilon} \Psi^\mu , \delta \psi^\mu = -i
\rho^\alpha \epsilon (\partial_\alpha X^\mu - \bar{\Psi}^\mu
\chi_\alpha ), \]
and
\begin{equation}
\delta e_\alpha = - 2 i \bar{\epsilon} \rho^\alpha \chi_\alpha ,
\delta \chi_\alpha = \nabla_\alpha \epsilon.
\end{equation}
The other transformations which leave this action invariant as listed
in referernces~\cite{bd11} and ~\cite{bd10}. Variation of the action
with respect to 
$e_\alpha$ leads to the vanishing of the energy momentum tensor while 
the variation with respect to $\chi^\alpha$ gives $J_\alpha = 0$. The postulates
made in equation (\ref{a5}) are now derived from a gauge principle. However, this
action is {\underline{not}} invariant under the four dimensional
space-time supersymmetry.

In four dimensions, the Green-Schwarz action for $N=1$ local
supersymmetry is worked out in references  \cite{bd5} and
\cite{bd13}.
\begin{equation}
S = \frac{1}{2\pi} \int d^2 \sigma \left ( \sqrt{g} g^{\alpha \beta}
\Pi_\alpha \cdot \Pi_\beta + 2 i\epsilon^{\alpha\beta}
\partial_\alpha X^\mu  \bar{\theta} \Gamma_\mu \partial_\beta 
\theta  \right ),\label{a6}
\end{equation}
where
\begin{equation}
\Pi_\alpha^\mu = \partial_\alpha X^\mu - i \bar{\theta} \Gamma^\mu
\partial_\alpha \theta.
\end{equation}
and where the $\theta$ is a  genuine space-time
fermion fields rather than spacetime vectors. In our case they are
$D=4$, four component spinors of $SO(3,1)$, constructed like
the $\Psi^\mu$.  $\Gamma^\mu$ are the four dimensional Dirac gamma
matrices. This action is invariant under global supersymmetry
\[ \delta \theta = \epsilon \]
and
\begin{equation}
\delta X^\mu = i \bar{\epsilon} \Gamma^\mu \epsilon
\end{equation}
provided
\begin{equation}
\Gamma_\mu \Psi_{[1} \bar{\psi}_2 \Gamma^\mu \Psi_{3]} = 0
\end{equation}
and this is satisfied for our  Majorana or Weyl spinors in four dimensions
\cite{bd5}. 

Consider a supersymmetric parameter $\kappa^{ a \alpha}$ where
$ \alpha$ represents a two dimensional vector index and `$a$' is
a spinoal index. It has been
shown in detail in reference \cite{bd5}, that the action is invariant
under $\kappa$ transformation wherever global $\epsilon$ symmetry
invariance exists as in  four cases listed in reference [9].

The transformations are
\begin{equation}
\delta \theta = 2 i \Gamma \cdot \Pi_\alpha \kappa^{\alpha}
\end{equation}
and
\begin{equation}
\delta \chi^\mu = i \bar{\theta} \Gamma^\mu \delta \theta.
\end{equation}
Thus the action (\ref{a6}) is obviously spacetime supersymmetric in four
dimensions.

The major defect with this GS action is that the naive covariant
quantisation procedure does not work. Only in the light cone gauge,
things are much simpler and the theory can be quantised. If one wishes 
to proceed with a covariant formulation, one has no other choice but
to implement the NS-R scheme with GSO projection which is simple and
elegant and also equivalent. It can be
covariantly quantised and the critical central charge can be calculated. We
follow this scheme in the following sections.

\section{Quantization}
As usual the theory is quantized  ($\alpha_o^{\mu} = p^{\mu}$), with 
\begin{equation}
X^{\mu}=x^{\mu}+p^{\mu}\tau +i\sum_{n\neq 0}\frac{1}{n}\alpha^{\mu}_n
e^{-i n\tau} cos(n\sigma), \nonumber
\end{equation}
or
\begin{equation}
\partial_{\pm}X^{\mu}=\frac{1}{2}\sum_{-\infty}^{+\infty}
\alpha_n^{\mu}\; e^{-in(\tau\pm\sigma)}
\end{equation}
and
\begin{equation}
[\alpha_m^{\mu},\alpha_n^{\nu}]=m\; \delta_{m,-n}\;
\eta^{\mu \nu}.
\end{equation}

We mention in passing that for closed strings
\begin{equation}
\partial_{-}X^R_{\mu}=\sum_{-\infty}^{+\infty}\alpha^{\mu}_n
~~e^{-2in(\sigma-\tau)}
\end{equation}
and
\begin{equation}
\partial_{+}X^L_{\mu}=\sum_{-\infty}^{+\infty}\tilde{\alpha}^{\mu}_n
~~e^{-2in(\sigma+\tau)}
\end{equation}
The entire algebra can be made applicable to closed string as well.

We first choose the Neveu-Schwarz (NS) \cite{bd4}
boundary condition. Then the mode expansions of the fermions are
\begin{equation}
\psi_{\pm}^{\mu,j}(\sigma,\tau)=\frac{1}{\sqrt 2}
\sum_{r\in Z+\frac{1}{2}}b_r^{\mu,\;j} ~~e^{-ir (\tau\pm\sigma)}
\end{equation}
and
\begin{equation}
\phi_{\pm}^{\mu,k}(\sigma,\tau)=\frac{1}{\sqrt 2}
\sum_{r\in Z+\frac{1}{2}}b_r^{\prime\,\mu,\;k}~~e^{-ir (\tau \pm \sigma)}
\end{equation}
with real quanta ${\bf b}^\prime$, the quantisation is like
\[ \phi^\mu_+ (\sigma) = \frac{i}{\sqrt{2}} \sum_{r > 0} \left ( 
{\bf b}^{\prime \mu}_r ~e^{-i r \sigma} - {\bf b}_r^{+ \prime \mu} ~
e^{+ i r \sigma} 
\right ) \]
so that ~~$\{{\bf b}^{+ \prime \mu}_r , {\bf b}^{\prime \nu}_{s} \} =
\eta^{\mu \nu} \delta_{r,s}$ ~~and similar relations follow. We also have
\begin{equation}
\Psi_{\pm}^{\mu}(\sigma,\tau)=\frac{1}{\sqrt 2}
\sum_{r\in Z+\frac{1}{2}}B_r^\mu~ e^{-ir (\tau\pm\sigma)}
\end{equation}
\begin{equation}
 with ~~b_{r,j}^\mu = e_{\psi,j} B_r^\mu ,~~~~  b_{r,k}^\prime \mu = i
e_{\phi,k} B^{\mu}_r, ~~~~
 B^\mu_r =  e^j_\psi b_{r,j}^{\mu} + i e_\phi^k b_{r,k}^{\prime\mu}. 
\end{equation}
The sum is over all the half-integer modes. The anticommutation relations are
\begin{equation}
\{b_r^{\mu,j}, b_s^{\nu, j^{\prime}}\}=\eta^{\mu \nu}\;\delta_{j,j^{\prime}}\; 
\delta_{r,-s},
\end{equation}
\begin{equation}
\{b_r^{\prime\,\mu,k}, b_s^{\prime\,\nu, k^{\prime}}\} = 
\{ {\bf b}^{\prime \mu, k}_r , {\bf b}_s^{\prime \nu , k^\prime} \} =
\eta^{\mu \nu}\;\delta_{k,k^{\prime}}\;\delta_{r,-s}
\end{equation}
and
\begin{equation}
\{B_r^{\mu}, B_s^{\nu}\}=\eta^{\mu \nu}\; \delta_{r,-s}\;.
\end{equation}

\section{Virasoro Algebra}
Virasoro generators \cite{bd6} are given by the modes of the
energy momentum tensor $T_{++}$ and the Noether current $J_+$,
\begin{equation}
L_m^M=\frac{1}{\pi}\int_{-\pi}^{+\pi} d\sigma\; e^{im\sigma}\; T_{++},
\end{equation}
\begin{equation}
and ~~~~~~G_r^M=\frac{\sqrt 2}{\pi}\int_{-\pi}^{+\pi} d\sigma\;
e^{ir\sigma}\; J_{+}.
\end{equation}
`$ M $' stands for matter. In terms of creation and annihilation
operators
\begin{equation}
L_m^M=L_m^{(\alpha)}+L_m^{(b)}+L_m^{(b')},
\end{equation}
where
\begin{equation}
L_m^{(\alpha)}=\frac{1}{2}\sum_{n=-\infty}^{\infty}:
\alpha_{-n}\cdot\alpha_{m+n}:,
\end{equation}

\begin{equation}
L_m^{(b)}=\frac{1}{2}\sum_{r=-\infty}^{\infty}
(r+\frac{1}{2}m) :b_{-r}\cdot b_{m+r}:,
\end{equation}

\begin{equation}
and ~~~L_m^{(b')}=\frac{1}{2}\sum_{r=-\infty}^{\infty}
(r+\frac{1}{2}m):b'_{-r}\cdot b'_{m+r}:.
\end{equation}

In each case normal ordering is required. The single dot
implies the sum over all qualifying indices. 
The current generator is
\begin{equation}
G_r=\sum_{n=\infty} \alpha_n\cdot B_{r+n}.
\end{equation}
Following from the above equations,  the Virasoro algebra is
\begin{equation}
[L_m^M, L_n^M]=(m-n) L_{m+n}^M+A(m)\;\delta_{m,-n}
\end{equation}
Using the relations
\begin{equation}
\left [L^M_m,\alpha_n^{\mu} \right ]=-n~\alpha^{\mu}_{n+m},
\end{equation}
\begin{equation}
and~~~~\left [L^M_m,B_n^{\mu} \right ]=-(n+\frac{m}{2})B^{\mu}_{n+m}.
\end{equation}
we get
\begin{equation}
[L_m^M, G^M_r]=\left (\frac{1}{2}m-r\right ) G_{m+r}^M.
\end{equation}
The anticommutator $\{G^M_r, G^M_s\}$ is obtained by the use of the 
Jacobi Identity
\begin{equation}
[\{G^M_r,G^M_s\},L_m^M]+\{[L_m^M,G^M_r],G^M_s\}+\{
[L_m^M,G^M_s],G^M_r\}=0,
\end{equation}
which implies, consistent with equations (61) and (64),
\begin{equation}
\{G^M_r,G^M_s\}=2 L_{r+s}^M+B(r)\delta_{r,-s}
\end{equation}
$A(m)$ and $B(r) $ are normal ordering anomalies. Also note
that $L_{m}^{\dagger} = L_{-m}$ and $G_{r}^{\dagger} = G_{-r}$. Taking the 
vacuum expectation value in the Fock ground state $|0,0\rangle $ 
with four momentum $ p^{\mu}=0$ of the commutator $[L_1,L_{-1}]$
and $[L_2, L_{-2}]$, it is easily found that
\begin{equation}
A(m)=\frac{26}{12}(m^3-m) = \frac{C}{12}(m^{3}-m),
\end{equation}
and using the Jacobi Identity

\begin{equation}
B(r)=\frac{A(2r)}{2r}
\end{equation}
\begin{equation}
and ~~~~~~B(r) = \frac{26}{3}\left (r^2-\frac{1}{4}\right ) =
\frac{C}{3} 
\left ( r^2 - \frac{1}{4} \right )
\end{equation}

The central charge $C=26$. This is what is expected.
Each bosonic coordinate contribute 1 and each fermionic
ones contribute $1/2$, so that the total central charge is +26.
 The central charge can also be calculated from the leading divergency 
 of the  vaccum expection value of the product of two energy momentum
 tensors at two world sheet points $x$ and $w$. Since
\[ \langle X^\mu (z) , X^\nu (w) \rangle \sim \eta^{\mu \nu} \log
(z-w) \],
\[ \langle \psi^{\mu j} (z), \psi^\nu_{j^\prime} (w) \rangle \sim
\eta^{\mu \nu} \delta^j_{j^\prime} (z-w)^{-1} \]
and
\[ \langle \phi^{\mu k} (z), \phi^\nu_{k^\prime} (w) \rangle \sim \eta^{\mu \nu}
\delta^k_{k^\prime} (z-w)^{-1} ,\]
we deduce that,  $2\langle T_+ (z) T_+ (w) \rangle \sim C(z-w)^{-4} +
\cdots $ where $C = \eta_\mu^\mu + \frac{1}{2} \eta_\mu^\mu \delta^j_j 
+ \frac{1}{2} \eta_\mu^\mu \delta_k^k = 26. $

\section{Bosonic States and elimination of Ghosts}
A physical bosonic state $\Phi$ can be convinently constructed by operating the
generators $L$'s and $G$'s on the vacuum. They satisfy

\begin{equation}
( L_0^M - 1 ) \mid \Phi \rangle = 0,\label{a}
\end{equation}
\begin{equation}
L_m^M \mid \Phi \rangle = 0, \; \; \; m > 0,
\end{equation}
\begin{equation}
and~~~~~~G_r^M \mid \Phi \rangle = 0, \; \; \; r > 0
\end{equation}

These physical state conditions enable to exclude the time-like quanta from the
physical spectrum. Before doing this rigorously, we first follow the
qualitative arguments of Green, Schwarz and Witten
\cite{bd5}. Specialising to a rest frame, we 
write the physical condition for $L_m$ as
\begin{equation}
\frac{1}{2} p^{0} \alpha_{m}^{0} \mid \Phi \rangle \; + \; {\rm
  (terms~quadratic~in~osillators)} \mid \Phi \rangle = 0.
\end{equation}
In this frame, the physical states are generated effectively by the
space components of the oscillators only; so that
$\alpha_{m}^{0}\mid\Phi \rangle = 0$ following from the constraint
that the energy momentum tensor vanishes. Using the condition for the
generator $G_r$,
\begin{equation}
 \left [ G_{r}^M, \alpha_{m}^{0} \right ] \mid \Phi \rangle = m
~B_{m+r}^{0} \mid \Phi \rangle = 0  .
\end{equation}
Since $B^0_{m+r} \mid \Phi \rangle = 0$, ~~~~~$B_r^0 \mid \Phi \rangle = 0$~~~~~
as $B^0_{m+r}$ anticommutes with $B^0_r$.
Specialising to $\mu = 0$,  we arrive at the  initutive result from
equation (53)
\begin{equation}
b_r^{0,j} \mid \Phi \rangle = 0
\end{equation}
and
\begin{equation}
b_r^{\prime 0,k} \mid \Phi \rangle = 0.
\end{equation}
Thus the vanishing 
of the energy-momentym tensor and the current excludes all the time
like quanta from the physical space. No negative norm state is
expected to  show up in the physical spectrum.

Let us make a detailed investigation to ensure that there are no
negative norm physical states. We shall do this by constructing the
zero norm states or the `null' physical states. Due to the  GSO condition,
which we shall study later, the physical states will be obtained by
operation of the product of even number of G's. 
We construct generic states $|\tilde{\chi}\rangle$ such that 
$L_o |\tilde{\chi}\rangle =(1-n)|\tilde{\chi}\rangle$.~ n=0~ gives 
the tachyon $|\tilde{\chi}_o\rangle$. The next higher state is 
$|\tilde{\chi}_{1/2}\rangle$ is another fermionic tachyon 
and is projected out by GSO. So the next lowest allowed state is

\[ \mid \Psi \rangle = L_{-1} \mid \tilde{\chi}_{1} \rangle +\lambda~ G_{-1/2}
G_{-1/2} \mid \tilde{\chi}_{1} \rangle \]
But $ G_{-1/2} G_{-1/2} = \frac{1}{2} \{  G_{-1/2}, G_{-1/2} \} =
L_{-1}$. Without loss of generality, the state is
\begin{equation}
\mid \Psi \rangle = L_{-1} \mid \tilde{\chi} \rangle
\end{equation}
This state to be physical, it must satisfy $( L_{0} - 1) \mid \Psi
\rangle = 0$ which is true if $L_{0} \mid \tilde{\chi} \rangle = 0$. The
norm $\langle \Psi \mid \Psi \rangle = \langle \tilde{\chi}\mid L_{1} L_{-1} \mid
\tilde{\chi} \rangle = 2 \langle \tilde{\chi} \mid L_{0} \mid
\tilde{\chi} \rangle = 0$. Let us consider the next higher allowed mass state
\begin{eqnarray}
 \mid \Psi \rangle && = L_{-2} \mid \tilde{\chi}_{2} \rangle +a~ L_{-1}^{2} \mid
\chi_{2} \rangle + b ( G_{-3/2} G_{-1/2} + \lambda G_{-1/2} G_{-3/2} )
\nonumber \\
 && \mid \tilde{\chi}_{2} \rangle +c~ G_{-1/2} G_{-1/2}G_{-1/2}G_{-1/2} \mid
\tilde{\chi}_{2} \rangle  + \cdots \nonumber
\end{eqnarray}
It can be shown that ~~$G_{-3/2} G_{-1/2} \mid \tilde{\chi_1} \rangle = (
\beta_{1} L_{-1}^{2} + \beta_{2} L_{-2}) \mid \tilde{\chi_1}
\rangle$.~~~~~ The coefficients $\beta_{1}$ and $\beta_{2}$ can be
calculated by evaluating ~~~~~$\left [ L_{1}, G_{-3/2} G_{-1/2} \right ]
\mid \tilde{\chi_1} \rangle$ ~~~and~~~ $\left [ L_{2}, G_{-3/2} G_{-1/2} \right ]
\mid \tilde{\chi_1} \rangle$.~~ $G_{-1/2}^{4}$ ~is proportional to
$L_{-1}^{2}$. So, in essence, we have the next excited state as
\begin{equation}
\mid \Psi \rangle = \left ( L_{-2} + \gamma L_{-1}^{2} \right ) \mid
\tilde{\chi}_1 \rangle.
\end{equation}
The condition $(L_{0} - 1)\mid\Psi\rangle=0$ is satisfied if $(L_{0} + 1) 
\mid \tilde{\chi_1} \rangle = 0$. Further the physical state condition
$L_{1} \mid \Psi \rangle = 0$ gives the value of $\gamma = 3/2$. The
norm is easily obtained as 
\begin{equation}
\langle \Psi \mid \Psi \rangle = \frac{1}{2} (C - 26).
\end{equation}
This is negative for $C < 26$ and vanishes for $C=26$. So the critical 
cenbtral charge is 26. It is easily checked that $L_{2} \mid \Psi
\rangle$ also vanishes for $C=26$.

To find the role of $b$ and $b^{\prime}$ modes, let us calculate the norm
of the following state with $p^{2} = 2$
\begin{eqnarray}
(L_{-2} + 3/2 L_{-1}^{2} ) \mid 0, p \rangle &=& \left (
L_{-2}^{(\alpha)} + \frac{3}{2} L_{-1}^{(\alpha)^{2}} \right ) \mid, 
0, p \rangle  \nonumber \\
&& + \left (  L_{-2}^{(b)} + \frac{3}{2} L_{-1}^{(b)^{2}}
\right ) \mid 0, p \rangle  + \left (  L_{-2}^{(b^{\prime})} +
  \frac{3}{2} L_{-1}^{(b^{\prime})^{2}} \right ) \mid 0, p \rangle. \nonumber
\end{eqnarray}
The norm of the first term is equal to $-11$ as calculated in reference
\cite{bd5}.

Noting that $L_{-1}^{(b)} \mid 0,p\rangle = L_{-1}^{(b^{\prime}}) \mid
0, p \rangle = 0$;~~~  $L_{-2}^{(b)} = \frac{1}{2} b_{-3/2} \cdot
b_{-1/2}$ ~and~ $ L_{-2}^{(b^{\prime})} = \frac{1}{2} b_{-3/2}^{\prime} \cdot 
b_{-1/2}^{\prime}$ ~~~the norms of the second and third terms are
~~$\frac{1}{4} (\delta_{\mu \mu} \delta_{j j}) = 6$ ~~and~~ $\frac{1}{4}
(\delta_{\mu \mu} \delta_{k k}) = 5$ ~~respectively. The norm of the
state given above is $-11 + 6 + 5 = 0$

The ghosts can be eliminated from the tree amplitudes in superstrings
in the RNS formulation. The proof given in reference 9, is easily
adapted for this string as well. We shall mention certain differences
in this superstring. The tree amplitude is 
\[ A_M = g^{M-2} \langle \phi_1 \mid V(2) \Delta V(3) \cdots \Delta
V(M-1) \mid \Phi_M \rangle . \]
$\mid \phi \rangle$, $V$ and $\Delta$ are suitably choosen to
eliminate ghosts. $V$, the vertex operator function should have
conformal weight 1. For physical states $ (L_0 - 1 \mid \Phi \rangle =
0$ is the wave equation, so that 
$\Delta = (L_0 - 1)^{-1}$ is the propagator. This is similar to the $F_1$ 
picture in the usual superstring theory.

We have to show that
\begin{equation}
\langle \tilde{\chi} \mid L_n V(2) \Delta V(3) \cdots \Delta (M-1)
\mid \Phi_M \rangle=0,
\end{equation}
where $\langle \tilde{\chi} \mid$ is any state satisfying $(L_0 - 1 +
n ) \mid \tilde{\chi} \rangle = 0$.

We need two identities
\begin{equation}
[ L_n - L_0 - n + 1 ] V = V [ L_n - L_0 + 1 ],
\end{equation}
\begin{equation}
and ~~~~[ L_n - L_0 + 1 ] \frac{1}{L_0 - 1} = \frac{1}{L_0 + n - 1} [ L_n -
L_0 - n + 1 ] .
\end{equation}
 $L_n$ which is in effect the same as $L_n - L_0 - n + 
1$, can be pused to the right till the factor at the end. $(L_n - L_0 + 
1) \mid \Phi_M \rangle$  is zero.

The $F_2$ like picture is very illiminating from the point of view of
this superstring. As observed already constructing a new state
$\tilde{\Phi}$ through $\Phi = G_{-1/2} \tilde{\Phi}$, we obtain $(L_0
- 1/2) \mid \tilde{\Phi} \rangle = 0 $. If we
supplement this with a physical state like consition $G_{1/2} \mid
\tilde{\Phi} \rangle = 0$, we deduce that $G_{1/2} \mid \Phi \rangle = \mid
\tilde{\Phi} \rangle$ and $G_r \mid \Phi \rangle = G_r \mid
\tilde{\Phi} \rangle = 0 $ for $r \geq \frac{3}{2}$. As a consequence, 
the above tree amplitude of the $F_1$ like picture becomes
\begin{eqnarray}
 A_M &&  = g^{M - 2} \langle \tilde{\Phi} \mid G_{1/2} V(2) \Delta V(3) \cdots
\Delta V(M-1) \mid \Phi_M \rangle \nonumber \\
&&  = g^{M - 2} \langle \tilde{\Phi} \mid  V(2) \tilde{\Delta} V(3) \cdots
\tilde{\Delta} V(M-1) \mid \tilde{\Phi}_M \rangle, \nonumber
\end{eqnarray}

where the $\tilde{\Delta} =  (L_0 - \frac{1}{2})^{-1}$ is the new
propagator.

Now we have to prove
\begin{equation}
\langle \tilde{\chi} \mid G_r V(2) \bar{\Delta} V(3) \cdots
\bar{\Delta} V(M-1) \mid \tilde{\Phi}_M \rangle = 0,
\end{equation}
where the state $\langle \tilde{\chi} \mid$ is such that it is
anihilated by $L_0 - \frac{1}{2} + r$. Now we have to move $G_r$ to
the right. We introduce an operator $W$ of conformal weight
$\frac{1}{2}$ such that$V = \{ G_r, W \}$ and $[ G_r , V] = [ L_{2r},
W ]$. However, the commutator can be replaced by
\begin{equation}
[ L_{2r} - L_0 - r + \frac{1}{2}] W - W(L_{2r} - L_0 + \frac{1}{2}) =
0.
\end{equation}
Thus $G_r$ is just switched over to the right of $V$. Using
\begin{equation}
G_r \frac{1}{L_0 - \frac{1}{2}} = \frac{1}{L_0 + r - \frac{1}{2}} G_r,
\end{equation}
we can bring past the $V$ and $\bar{\Delta}$'s till it anihilates
against $\mid \tilde{\Phi}_M \rangle$.

Eventhough, it can be easily shown that the ghosts are eliminated from 
the trees, it is less obvious for the loops. It is contended that, for 
the loops,
it should be done on a case to case basis.

\section{Conformal Ghosts}
For obtaining a zero central charge so that the anomalies
cancel out and natural ghosts are isolated, Faddeev-Popov 
(FP) ghosts \cite{bd7} will be introduced. The FP ghost action
is
\begin{equation}
S_{FP}=\frac{1}{\pi}\int (c^+\partial_- b_{++} + c^-
\partial_+b_{--})d^2 \sigma,
\end{equation}
where the ghost fields $b$ and $c$ satisfy the anticommutator
relations
\begin{equation}
\{b_{++}(\sigma,\tau), c^+(\sigma^{\prime},\tau)\}
=2 \pi\; \delta(\sigma-\sigma^{\prime})
\end{equation}

\begin{equation}
and ~~~~\{b_{--}(\sigma,\tau), c^-(\sigma^{\prime},\tau)\}
=2 \pi\; \delta(\sigma-\sigma^{\prime})
\end{equation}
and are quantized with the mode expansions
\begin{equation}
c^{\pm}=\sum_{-\infty}^{\infty}c_n\; e^{-in(\tau\pm\sigma)}
\end{equation}
and
\begin{equation}
b_{\pm \pm}=\sum_{-\infty}^{\infty}b_n\; e^{-in(\tau\pm\sigma)}
\end{equation}
The canonical anticommutator relations for $c_n$'s and
$b_n$'s are
\begin{equation}
\{c_m,b_n\}=\delta_{m,-n},~~~~~~\{c_m,c_n\}=\{b_m,b_n\}=0
\end{equation}

Deriving the energy momentum tensor from the action and making
the mode expansion, the Virasoro generators for the ghosts (G)
are
\begin{equation}
L_m^G=\sum_{n=-\infty}^{\infty}(m-n)\;b_{m+n}\; c_{-n}- a\; \delta_{m,-n}
\end{equation}
where $a$ is the normal ordering constant. These generators 
satisfy the algebra
\begin{equation}
[L_m^G,L_n^G]=(m-n)\;L_{m+n}^G+A^G(m)\; \delta_{m,-n}
\end{equation}
The anomaly term is deduced as before and give
\begin{equation}
A^G(m)=\frac{1}{6}(m-13m^3)+2a\;m.
\end{equation}
With $a=1$, this anomaly term becomes

\begin{equation}
A^G(m)=-\frac{26}{12}(m^3-m),~~~~~~~
B^{G} (r) = - \frac{26}{3} \left ( r^{2} - \frac{1}{4} \right )
\end{equation}

The central charge is $-26$ and cancels the normal ordder $A(m)$ and
$B(r)$ of the $L$ and $G$ generators.

Since $G_r^{gh}$, the ghost current gauge factor has conformal weight
$\frac{3}{2}$, 

\begin{equation}
[L_m^G,G_r^{gh}]=(m/2-r)G^{gh}_{m+r}.
\end{equation}

 From the Jacobi Identity
\begin{equation}
\{ G_{r}^{gh}, G_{s}^{gh} \} = 2 L_{r+s}^{G} + \delta_{r,-s} B^{G}(r).
\end{equation}
It immedicately follows that 
\begin{equation}
G_{r}^{gh^{2}} = L^{G}_{2r}.
\end{equation}
 In practice, the products of even number of 
$G_{r}^{gh}$'s occur in calculations and they can be evaluated in
terms of $L_{2r}^{G}$'s.

The total current generator is 
\begin{equation}
G_r=G_r^M+G^{gh}_r,
\end{equation}
and we have the anomaly free Super Virasoro algebra,

\begin{equation}
[L_m,L_n]=(m-n)L_{m+n},
\end{equation}

\begin{equation}
[L_m,G_r]=(m/2-r)G_{r+m}
\end{equation}
and
\begin{equation}
[G_r,G_s]=2L_{r+s}
\end{equation}
 Thus from the usual conformal field theory we have 
obtained the algebra of a superconformal field theory. This is the novelty 
of the proposed string.The BRST \cite{bd8} charge operator taking the case of
constraints (\ref{a}) is 
\begin{equation}
Q_{1}=\sum_{-\infty}^{\infty}L_{-m}^M\;c_m -\frac{1}{2}
\sum_{-\infty}^{\infty}(m-n) :c_{-m}\; c_{-n}\; b_{m+n} :
-a\; c_0.
\end{equation}
and is nilpotent for $a=1$. This is the open bosonic string.

The absence of the need to use superconformal ghosts may appear
puzzling. But we note that the light cone gauge is ghost
free. Dropping the helicity suffixes, the light cone vectors
\begin{equation}
\psi_j^\pm = \frac{1}{\sqrt{2}} (\psi_j^0 \pm \psi_j^3 )~~~~~~
\phi_k^\pm = \frac{1}{\sqrt{2}} ( \phi_k^0 \pm \phi_k^3 ),
\end{equation}
have the anticommutators with negative signs.
\begin{equation}
\{ \psi_j^+ (\sigma) , \psi^-_{j^\prime} (\sigma^\prime) \} = - 
\delta_{jj^\prime} \pi \delta(\sigma-\sigma^\prime) ; \{\phi_k^+ 
(\sigma) , \phi_{k^\prime}^- (\sigma^\prime) \} = - \delta_{kk^\prime} \pi \delta
(\sigma-\sigma^\prime).
\end{equation}
The total ghost energy momentum tensor comes from the $(0,3)$
coordinates and is given by
\begin{equation}
T^{gh} (z) = \frac{i}{2} \left ( \psi^{0j} \delta_z \psi_{0j} +
\psi^{3j} \delta_z \psi_{3j} \right ) + \frac{i}{2} \left (\phi^{0k}
\delta_z \phi_{0k} + \phi^{3k} \delta_z \phi_{3k} \right ).
\end{equation}
But the correlation function
\begin{equation}
\langle T^{gh} (z) , T^{gh} (\omega) \rangle = \frac{11}{2}
\frac{1}{(z-\omega)^4} + \cdots
\end{equation}
Thus the contribution of these ghosts to the central charge 26 is
11. The remaining charge 15 is the same as for a normal ten
dimensional superstring. See appendix for a study of equivalence with
FP action. With introduction of FP($\beta ,\gamma$) ghosts, the BRST 
charge changes by
\begin{equation}
Q'_{BRST}=\sum G_{-r}~\gamma_r -~\sum \gamma_{-r}~\gamma_{-s}~b_{r+s},
\end{equation}
in the NS sector and similarly for the R sector with integral r. This
takes care of the current constraints.

\section{Superconformal ghosts}

The fermionic light cone part of the action can be written as
\begin{equation}
S_F^{l.c}= -\frac{i}{\pi} \int d^2\sigma~\bar{\Psi}^+\rho^{\alpha}
\partial_{\alpha}\Psi^-
\end{equation}
We supplement this Hilbert space by a space where the fermi particles 
satisfy bose statistics. This is done here by letting $\bar{\Psi}^+$~~and ~~ 
$ \Psi^-$ ~~depend on Grassman variable $\bar{\theta}$~~ and
~~$\theta$  so that the action
\begin{equation}
S_F^{l.c}=- \frac{i}{\pi} \int d^2\sigma~\int d\bar{\theta}\int d\theta~
\bar{\Psi}^+(\bar{\theta})~\rho^{\alpha}\partial_{\alpha}
\Psi^-(\theta).
\end{equation}
Expanding the fields as 
\begin{equation}
\bar{\Psi(\bar{\theta} )}^+= \cdots + \bar{\theta}\bar{\gamma} +\cdots
\end{equation}
and
\begin{equation}
+2~i~\rho^{\alpha}\Psi^-(\theta) = \cdots  + \theta\beta^{\alpha}+\cdots
\end{equation}

We obtain the Superconformal ghost action~\cite{bd5} as
\begin{equation}
S_F^{l.c}= -\frac{1}{2\pi} \int
d^2\sigma~\bar{\gamma}\partial_{\alpha}\beta_{\alpha}
\end{equation}
The rest is standard. The wave equations are ~~$\partial\gamma
=\partial\beta=0$~~. The energy momentum tensor is
\begin{equation}
T_{++}= -\frac{1}{4}\gamma~\partial_+\beta -\frac{3}{4}\beta\partial_+\gamma.
\end{equation}
In the quantised form
\begin{eqnarray}
\gamma(\tau) &=&\frac{1}{\sqrt{2}} \sum\gamma_{n(r)}~e^{-2 i\tau n(r)}\nonumber\\
\beta(\tau) &=&\frac{1}{\sqrt{2}} \sum\beta_{n(r)}~e^{-2 i\tau n(r)}\nonumber
\end{eqnarray}
for integral ~ n~  or half integral ~r~. The nonvanishing relation is
$[\gamma_m, \beta_n ]= \delta_{m,-n}$. $L_m^{l.c.}$ which is contained 
in $L_n^M$ is now (Ramond)
\begin{equation}
L_m^{l.c.} = \sum_n ~(n + m/2) ~~:\beta_{m-n}\gamma_n~:\label{b}
\end{equation}
In component notations
\begin{equation}
L_m^{l.c.} = \sum_n ~(n + m/2) \left ( \beta^j_{m-n}\gamma_{n,j}~+~
\beta^k_{m-n}\gamma_{n,k}\right ).
\end{equation}
The conformal dimension of $\gamma$ is `-1/2' and $\beta$ is `3/2' as can be 
deduced from (\ref{b}),
\begin{equation}
[ L_m^M, \gamma_n ]=(-3/2 - n )\gamma_{m+n}\label{c}
\end{equation}
\begin{equation}
[ L_m^M, \beta_n ]=(3/2 - n )\beta_{m+n}
\end{equation}
The part of the BRST   charge which takes care of the current
generator constraints(\ref{a}), for instance, is
\begin{equation}
Q'=\sum_r~G_r\gamma_{-r}~-~ \sum_{rs}~\gamma_{-s}\gamma_{-r}~b_{s+r}.
\end{equation}
Similarly for Ramond sector.\\
The total BRST charge is then
\begin{equation}
Q_{BRST} = Q_1 + Q'
\end{equation}
With ~~$\{ Q_1, Q_1 \}$=0 , ~~it can be directly checked that ~~$\{ Q', Q'\} + 2
\{ Q_1, Q'\}$=0~~ by using the Virasoro algebra, equation (\ref{c}) and the 
following results obtained by Fourier transformation, integration by parts and 
field equation
\begin{equation}
\sum_r~\sum_s~\gamma_s~\gamma_r~\delta{r,-s} =
\sum_r~\sum_s~r^2~\gamma_s~\gamma_r~\delta{r,-s}=0.
\end{equation}
Thus 
\begin{equation}
Q_{BRST}= 0.
\end{equation}
The nilpotency of ~~$Q_{BRST}$~~ proves that the model is unitary and free 
from anomalies and ghosts.
\section{Fermionic States}
The above deductions can be repeated for Ramond sector~\cite{ramond}. 
We write the main equations. The mode expansion for the fermions are
\begin{equation}
\psi_{\pm}^{\mu,j}(\sigma ,\tau) = \frac{1}{\sqrt 2}
\sum_{-\infty}^{\infty} d_m^{\mu,j} e^{-im(\tau\pm\sigma)}
\end{equation}
\begin{equation}
\phi_{\pm}^{\mu,j}(\sigma ,\tau) = \frac{1}{\sqrt 2}
\sum_{-\infty}^{\infty} d_m^{'\mu,j} e^{-im(\tau\pm\sigma)}
\end{equation}

The generators of the Virasoro operators are
\begin{equation}
L_m^M = L_m^{(\alpha)} + L_m^{(d)} + L_m^{(d')}
\end{equation}
\begin{equation}
L_m^{(d)} = \frac{1}{2}\sum_{n=-\infty}^{\infty}(n+\frac{1}{2} m)
: d_{-n}\cdot d_{m+n} :
\end{equation}
\begin{equation}
L_m^{(d')} = \frac{1}{2}\sum_{n=-\infty}^{\infty}(n+\frac{1}{2} m)
: d'_{-n}\cdot d'_{m+n} :
\end{equation}
and the fermionic current generator is
\begin{equation}
F_m^M  = \sum_{-\infty}^{\infty}\alpha_{-n}\cdot D_{n+m}\label{d}
\end{equation}
The Ramond sector Virasoro algebra is the 
same as the NS-sector with the replacement of
G's by F's. It is necessary to define $L_o$ suitably to keep the anomaly 
equations
 the same~\cite{bd5}.

In this Ramond sector, a physical state $\mid \Phi \rangle$ should
satisfy
\begin{equation}
F_{n} \mid \Phi \rangle = L_{n} \mid \Phi \rangle = 0 \; \; \; {\rm
  for} \; \; \; n>0
\end{equation}

The normal order anomaly constant in the anticommutators of the Ramond 
current generators has to be evaluated with care, beacuse the
defination of $F_{0}$ does not have a normal ordering ambiguity. So
$F_{0}^{2} = L_{0}$. Using commutation relation for the $F$ and the 
Jacobi Identity we get
\[ \{ F_{r}, F_{-r} \} = \frac{2}{r} \{ [ L_{r}, F_{0} ], F_{-r}
\} = 2 L_{0} + \frac{4}{r} A(r) \]
So
\begin{equation}
B(r) = \frac{4}{r} A(r)
\end{equation}
\begin{equation}
B(r) = \frac{C}{3} (r^{2} - 1), \; \; \; \; r \neq 0
\end{equation}
A physical state in the fermionic sector satisfies
\begin{equation}
( L_{0} - 1 ) \mid \Psi \rangle = 0
\end{equation}
It follows that
\[ (F^{2}_{0} - 1 ) \mid \Psi \rangle  = (F_{0} - 1) (F_{0} + 1) \mid \Psi
\rangle = 0 \]

The construction of `null' physical states becomes much simpler
beacuse all $F_{-m}$ terms can be assigned to $L_{-m}$ terms by the
commutation ruation $F_{-m} = 2 [ F_{0}, L_{-m} ]/m$ and $F_{0}$ has
eigen values which are roots of eigen values of $L_{0}$ acting on the
generic states or states constructed out of the generic states. Thus 
the zero mass null physical state with $L_{0} \mid \tilde{\chi}
\rangle = F_{0}^{2} \mid \tilde{\chi} \rangle = 0$ is simply
\begin{equation}
\mid \Psi \rangle = L_{-1} \mid \tilde{\chi} \rangle
\end{equation}
with $L_{1} \mid \Psi \rangle = F_{1} \mid \Psi \rangle = 0$. The next
excited state with $(L_{0} + 1) \mid \tilde{\chi} \rangle $ becomes
the same as in the bosonic sector. Obtained from the condition 
$L_{1} \mid \Psi \rangle = 0$,
\[ \mid \Psi \rangle = (L_{-2} + \frac{3}{2} L_{-1}^{2}) \mid
\tilde{\chi} \rangle \]
The norm $\langle \Psi  \mid \Psi \rangle = (C - 26)/2$ and vanishes
for $C=26$. It is easy to check that all physical state conditions are 
satisfied. $F_{1} \mid \Psi \rangle = 2$ $[ L_{1}, F_{0} ] \mid \Psi \rangle = 0$
since $L_{1} \mid \Psi \rangle = 0$ and $F_{0} \mid \Psi \rangle =
\mid \Psi \rangle $, $L_{2} \mid \Psi \rangle = F_{1} F_{1} \mid \Psi
\rangle = 0$ and $F_{2} \mid \Psi \rangle = [ L_{2}, F_{0} ] \mid \Psi 
\rangle = 0$. For $C=26$, there are no negative norm states in the
Ramond sector as well.

\begin{equation}
\left [ L_{m}^{G}, F_{n}^{gh} \right ] = \left ( \frac{m}{2} -n \right 
) F_{m+n}^{gh}
\end{equation}
and 
\begin{equation}
\{ F_n^{gh}, F_{m}^{gh} \} = 2 L_{n+m}^{G} + B(r) \delta_{n, -m}.
\end{equation}

The ghost current anomaly constant is
$B^{G} (r) = - \frac{26}{3} (r^{2} - 1)$ from the Jacobi identity and cancels 
out the $B(r)$ of equation(\ref{d}). 

\section{The Mass Spectrum}
The ghosts are not coupled to the physical states.
Therefore the latter must be of the form (up to null state)\cite{bd9}.
\begin{equation}
|\{n\}\; p\rangle_M \otimes\; c_1|0\rangle_G.\label{eq64}
\end{equation}
$|\{n\}\; p\rangle_M$ denotes the occupation numbers and momentum of 
the physical matter states. The operator $c_1$ lowers the
energy of the state by one unit and is necessary for BRST
invariance. The ghost excitation is responsible for lowering
the ground state energy which produces the shiftable tachyon ($F_2$ picture). 
\begin{equation}
(L_0^M-1)\;|phys\rangle=0.
\end{equation}
Therefore, the mass shell condition is
\begin{equation}
\alpha^{\prime} M^2 = N^B+N^F_{NS}-1 ~~~~~~~~~~~~~~~~~~~: ~~~NS
\end{equation}
\begin{equation}
\alpha^{\prime} M^2 = N^B + N_R^F - 1 ~~~~~~~~~~~~~~~~~~:~~ R
\end{equation}
where 
\begin{equation}
N^B=\sum_{m=1}^{\infty}\alpha_{-m}\; \alpha_m
\end{equation}
and
\[
N^F_{NS}=\sum_{r=1/2}^{\infty}r\;(b_{-r}\;\cdot  
b_r+b'_{-r}\;\cdot b'_r)~~~~~~~~~~~~~~~~~~~~~~~~ : NS
\]
\begin{equation}
N_R^F = \sum_{m=1}^{\infty} m(d_{-m} \cdot d_m + d^{\prime}_{-m} \cdot 
d^{\prime}_{m} ) ~~~~~~~~~~~~~~~~~~~~~~~~: ~~R 
\end{equation}
In general, $\alpha^{\prime} M^2 = - 1, - \frac{1}{2}, 0, \frac{1}{2},
  1, \frac{3}{2}, \cdots$ in the N. S. sector. In the shifted Hilbert
  state $\alpha^\prime M^2 = -1/2 , 0, 1/2 , 1 , 3/2 , \cdots$ .

Due to the presence of Ramond and Neveu-Schwartz sectors with periodic and 
anti-periodic 
boundary conditions, we can effect a 
GSO projection ~\cite{bd10} on the mass 
spectrum on the NSR model~\cite{bd13}. The desired projection is
\begin{equation}
G = \frac{1}{2} (1 + (-1)^{F + F^{\prime}})
\end{equation}
where $ F = \sum b_{-r}\cdot b_r$ and    $F' =\sum b'_{-r}\cdot b'_r$ .
This will eliminate the half integral values 
from the mass spectrum by choosing G=1 including the tachyon at
$\alpha M^2 = - \frac{1}{2}$.

\section{Modular Invariance}

We shall follow the notation of Seiberg and Witten \cite{bd98} for the 
spin structures and the result, given by Kaku \cite{bd13}, that the
spin structure $\chi (--,\tau)$ for a single fermion is given by
\begin{equation}
\chi(--,\tau) =q^{-1/24} \;Tr\; q^{2\sum_n  n\;\psi_{-n} \psi_n}
=q^{-1/24} \prod_{n=1}^\infty (1+ q^{2n-1}) = 
\sqrt{\frac{\Theta_3(\tau)}{\eta(\tau)}}\;,
\end{equation}
where $\Theta$'s will be the Jacobi Theta functions 
\cite{bd98}, $q=e^{i\pi\tau}$ and 
$ \eta (\tau) (2\pi) = \Theta_1^{\prime 1/3}(\tau)$.  $\eta(\tau)$ is
the Dedekind eta function.

In the model there are three groups of oscillators, the four bosons, the twenty
four unprimed fermions and twenty primed fermions.There are six
constraint equations  $T_{++} = T_{--} = 0$ and
equations (\ref{b1}) and (\ref{b2}). In covariant formulation, the effective
number of physical modes is the total number of modes substracted by
the number of invariant constraints on the whole assembly. There are
two transverse bosonic and two `bosonic' ghost modes. For the group of 
twenty $(SO(5))$, there are eighteen and for the group of twenty four 
$(SO(6))$, there are twenty two independent fermionic modes. In all
there are forty of such fermions.

It is easy to construct the modular invariant partition function for the
two physical bosons, namely
\begin{equation}
{\cal P}_B(\tau) =(Im \;\tau)^{-2}\eta^{-2}(\tau){\bar \eta}^{-2}(\tau)\;,
\end{equation}
as given in reference [20]..

The partion function for the two bosonic ghosts can be calculated from 
the equations for the Hamiltonian given in reference \cite{bd5}.
\begin{equation}
{\cal P}_{\rm ghost} = \mid \frac{\Theta_2 (\tau)}{\eta(\tau)} \mid^4
\end{equation}

All that remains is to calculate the partition function of the forty
fermions. To obtain a physically viable partition, we observe that the 
initial internal symmetry of the 44 fermions was $SO(44)$. Then this  broke
down to four groups of 11 fermions, each group having the internal
symmetry $SO(11)$. $SO(11)$ can break to the maximal subgroups as  $SO(6) \times
SO(5)$. We would like to perform the  partition of  the forty fermions
in such a way 
that the $SO(6) \times SO(5)$ symmetry is obvious. The spinor
representation of $SO(6)$ has $2^3 = 8$ space time fermionic
modes. Let $b_{r,k}^i$ be the anihilation fermion quanta with the
$SO(6)$ deive spinor inded `i' running from 1 to 8 and the  $k$, a $SO(5)$
vector index running from 1 to 5. The $SO(6) \times SO(5)$ invariant
Hamiltonian in the NS sector is $H^{NS} = \sum_{k=1}^5 H_k^{NS}$, with 
$H_k^{NS} = \sum_r r b_{rk}^{\dagger i} b_{rk}^i - 8/48$. Further
$[H_k , H_{k^\prime} ] = 0$, there are five replicas of the group of
eight. We may consider five indentical boxes, each box containing
the eight fermions. The partition function of the forty fermions will be
the partition function of the eight fermions raised to the power of
five. 

The path integral functions of Seiberg and Witten \cite{bd98} for the 
eight fermions  in one box is 
\begin{equation}
A((--),\tau) = (\Theta_3 (\tau)/ \eta(\tau))^{4}\;,
\end{equation}
This is normalised to one. The other three spin structures are related
to the above function as given in \cite{bd98}.
\begin{equation}
A((+-),\tau) = A((--),\frac{\tau}{1+\tau}) = 
-(\Theta_2 (\tau)/ \eta(\tau))^{4}\;,
\end{equation}
and

\begin{equation}
A((-+),\tau) = A((+-), -\frac{1}{\tau}) =- (\Theta_4(\tau)/\eta(\tau))^{4}\;,
\end{equation}
\begin{equation}
A((++).\tau) =0\;.
\end{equation}
The sum of all spin structures is, therefore,
\[
A(\tau) = (\Theta_3 (\tau)/\eta(\tau))^{4} - 
(\Theta_2 (\tau)/\eta(\tau))^{4}-(\Theta_4(\tau)/\eta(\tau))^{4}
\]

Consider the amplitude
\begin{equation}
 A_N (\tau) = \eta^{4} (\tau) A(\tau) = \Theta_{3}^{4} (\tau) -
\Theta_2^{4} (\tau) - \Theta_4^{4} (\tau) 
\end{equation}
Since
\[ \Theta_3 (\tau + 1) = \Theta_4 (\tau), ~~~~~~~~~\Theta_2 (\tau + 1) = e^{i
  \pi/4} \Theta_2 (\tau), ~~~~~~~~~\Theta_4 (\tau + 1) = \Theta_3 (\tau), \] 
we have
\[ A_N (\tau + 1) = - A_N (\tau) \]
The invariance in $\tau \rightarrow - 1/ \tau$ is built in, in the ratios 
of the theta functions of each structure.

So the modular invariant partition function for all the  forty fermions is 
\begin{equation}
{\cal P}_F (\tau) = \mid A(\tau) \mid^{10}
\end{equation}

The total partition function for the model is the product integral
\[ Z = \int {\cal P}_B {\cal P}_{ghost} {\cal P}_F d \tau \]
We notice that due to the famous Jacobi relation, the $A(\tau)$, as
follows from  equation (143) is zero. So the modular invariant partition
function $Z$ vanishes. Thus the necessary condition for space-time
supersymmetry is satisfied. At the same time, the constant in the
power to which `q' is raised, is still, $-\frac{2}{24}$
$-\frac{2}{24}$ $-\frac{40}{48} = -1$, confirming the correctness of
our calculation.

\section{Supersymmetry Algebra}
One should 
examine  restictions imposed on the Fock space due to the
 a supersymmetric algebra. The supersymmetric
charge $Q$ should be such as to reproduce equations
\begin{equation}
\delta X^{\mu} = \left [ X^\mu , \bar{Q} \cdot \epsilon \right ]
= \bar{\epsilon} \cdot \Psi^{\mu}
\end{equation}
and
\begin{equation}
\delta \Psi^{\mu} = \left [ \Psi^{\mu}, \bar{Q} \cdot \epsilon \right
] = -i \rho^{\alpha} \partial_{\alpha} X^{\mu} \epsilon
\end{equation}
A simple inspection shows that
\begin{equation}
\bar{Q} = -\frac{i}{\pi} \int_{0}^{\pi} d \sigma \Psi_{\mu} \rho^{\alpha} \rho^0
\partial_{\alpha} X^{\mu}
\end{equation}
leading to 
\begin{equation}
Q^{\dagger} = -\frac{i}{\pi} \int_{0}^{\pi} d \sigma \Psi_{\mu}
\rho^{\alpha} 
\rho^0\partial_{\alpha} X^{\mu}
\end{equation}
and
\begin{equation}
Q = \frac{i}{\pi} \int_{0}^{\pi} \rho^0 \rho^{\alpha}{}^{\dagger} 
\partial_{\alpha} X^{\mu} \Psi_{\mu} d \sigma
\end{equation}
By a somewhat lengthy calculation it is deduced that 
\begin{equation}
\sum_{\alpha} \{ Q_{\alpha}^{\dagger}, Q_{\alpha} \} = 2H
\end{equation}
where $H$ is the Hamiltonian of the system. It follows that for the
any state $\mid \Phi_{0} \rangle$ in the Fock space 
\begin{equation}
\sum_{\alpha} \mid Q_{\alpha} \mid \phi_0 \rangle \mid^2 = 2 \langle
\phi_0 \mid H \mid \phi_0 \rangle \geq 0
\end{equation}
It appears essential that the tachyonic states disappear  from
the physical Fock space as the physical Hamiltonian is a sum of
squares of moduli of supersymmetric charges. The ground state is
massless. This can be understood as follows, $(L_0 -1)^{-1}$ being the 
scalar propagator the scalar tachyonic ground state energy is $\langle 
0 \mid (L_0^{NS} -1)^{-1} \mid 0 \rangle$. But there are two fermionic 
tachyons with same energy. Considering that there is a normal ordering 
negative sign for the loop fermions, the contribution is $-\langle 0
\mid (F_0 -1)^{-1} (F_0 +1)^{-1} \mid 0 \rangle = - \langle (L_0
-1)^{-1} \mid 0 \rangle$ exactly cancelling the scalar tachyonic
energy. This is as it should for supersymmetric theories.

Some admissible  Fock space states are

\[ NS~{\rm eigenstates:} \: \: \: \prod_{n, \mu} \prod_{m, \nu} \{
\alpha_{-n}^{\mu} \} \{ B_{-m}^{\nu} \} \mid 0 \rangle \]
\[  R~{\rm eigenstates:} \: \: \: \prod_{n, \mu} \prod_{m, 0} \{
\alpha_{-n}^{\mu} \} \{ D_{-m}^{\nu} \} \mid 0 \rangle u \]
$u$ is a spinor. GSO projection is implied for the N-S eigenstates.Let
us construct the zero mass modes. The tachyonic vacuum will be 
denoted by $|0\rangle$ and the zero mass ground state by $\mid \phi_0
\rangle$.

We start with the supergravity multiplet. The ground state
\begin{equation}
B^{\mu}_{-1/2} B^{\nu}_{-1/2} \mid 0 \rangle \epsilon_{\mu \nu}
\end{equation}
has zero mass. Due to the physical state conditions $G_{1/2} \mid
\phi_{0} \rangle = 0$
\begin{equation}
p^{\mu} \epsilon_{\mu \nu} = p^{\nu} \epsilon_{\mu \nu} = 0 
\end{equation}
It describes a massless antisymmetric tensor $A_{\mu \nu} = 1/2
(\epsilon_{\mu \nu} - \epsilon_{\nu \mu})$, which turns out to be a
pseudoscalar, a massless scalar $\epsilon_{\mu \mu}$ of spin $0$ and a 
massless symmetric terms  of spin 2: $\epsilon_{\mu \nu} = 1/2
(\epsilon_{\mu \nu} + \epsilon_{\nu \mu})$, which is traceless.

The other zero mass spinonial states are
\begin{equation}
\alpha_{-1}^{\mu} \mid 0 \rangle u_{1 \mu}
\end{equation}
\begin{equation}
D_{-1}^{\mu} \mid 0 \rangle u_{2 \mu}
\end{equation}
$u_{1 \mu}, u_{2 \mu}$ are spinor four vectors and are distinguished
by \cite{bd5},
\begin{equation}
\gamma_{5} u_{1 \mu} = u_{1 \mu}
\end{equation}
\begin{equation} 
\gamma_{5} u_{2 \mu} = - u_{2 \mu}
\end{equation} 
We shall consider them together as a four component spin vector
$u_{\mu}$. The condition $F_0 \mid \phi_0 \rangle = 0$, $F_1 \mid
\phi_0 \rangle = 0$, $L_1 \mid \phi_0 \rangle = 0$ lead to the
condition
\begin{equation}
\gamma \cdot p u_{\mu} = p^{\mu} u_{\mu} = \gamma^{\mu} u_{\mu} = 0
\end{equation}
This state  contains not only a spin $3/2$ but also a spin $1/2$
state. They can be projected out. The details have been given by GSO
in reference \cite{bd10}.

We now count the number of physical degrees of freedom :
\begin{tabbing}
fd;klfds;klfds;klfdsk;lfds;klfds;lkfds\=
klfdlkjfdskjlfdlkjfdslkjfdlkfdlkjfdklj\= lkfdsjlkfdsklfdskljfdsjlkf\= \kill
Graviton\> 2 degrees of freedom:\> $\rho^{a}_{\mu}$\>\\
Dilaton, $\epsilon_{\mu \mu}$, \> 1\> A\>\\
Antisymmetric tensor\> 1\> B\>\\
Spin 3/2\> 2\> $u_{\mu}$\>\\
spin 1/2\> 2\> u\>
\end{tabbing}

The numbers of the fermions and the bosons are equal. They can be grouped
together as the gravitational $(\rho^{a}_{\mu}, u_{\mu})$ and 
the matter $(A,B,u)$ multiplets.

The massless ground state vector is represented by
\begin{equation}
\alpha^{\mu}_{-1} \mid 0 \rangle \epsilon_{\mu} (p)
\end{equation}
Here, because of the $L_0$ condition, $p^2 \epsilon_\mu = 0$: The
constraint $ L_1 \mid \phi_0
\rangle = 0$ gives the Lorentz condition $p \cdot \epsilon = 0$. The
external photon polarisation vector can be subjected to an on shell
gauge transformation $\epsilon_{\mu}(p) \rightarrow \epsilon_{\mu} (p) 
+ \lambda p_{\mu}$. Therefore the state
\begin{equation} 
p_{\mu} \alpha^{\mu}_{-1} \mid 0 \rangle \lambda = L_{-1} \mid 0
\rangle \lambda
\end{equation}
decouples from the physical system. There are only two degrees of
freedom left. However, from the Ramond sector we have the spinor
\begin{equation}
p_{\mu} \alpha^{\mu}_{-1} \mid 0 \rangle u(p) = F_{-1} \mid 0 \rangle u(p)
\end{equation}
with $\gamma .p$ $u(p) = 0$ from the physical state condition. Further, 
as already noted, $\gamma_{5} u(p) = u(p)$. So the member of
the fermionic degrees of freedom is again two, just like the vector
boson. Thus for all the zero mass states the bosonic and the fermonic
degrees of freedom are equal. Before passing on to the next section,
we mention that space time supersymmetry  has been examined in a
standard like model in reference \cite{bd22}.

\section{Approach to standard like model}
In this section we start by constructing the gauge and matter fields
from the creation operators of the $NS$ and $R$ sectors.

Writing $b^{\mu}$ for the creation operator $b^{\mu}_{-1/2}$ of the NS
sector, the massless gauge bosons $A_{ij}^{\mu \nu}$ can be obtained
from the traceless field strength tensor of an adjoint representation,
\begin{eqnarray} 
F_{\mu \nu,ij} && = \epsilon_{\mu\nu\lambda\sigma}\left( b_{i}^{\lambda} 
b_{j}^{\sigma} - b_{j}^{\lambda}  b_{i}^{\sigma}  \right) \mid 0 \rangle 
\nonumber \\ 
&& = \partial_{\mu}  A_{\nu,ij} - \partial_{\nu} A_{\mu,ij} + g( 
A_{\mu,ik} A_{\nu,kj} - A_{\nu,ik} A_{\mu,kj} ) 
\end{eqnarray} 
When $i,j$ run from 1 to 11, we have the {\bf 55} of $SO(11)$; with 
$i,j$ running from 1 to 6 or 1 to 5, we get the {\bf 15} of $SO(6)$ or 
the {\bf 10} of $SO(5)$ respectively. Thus the SO(6)$\otimes$SO(5) 
symmetry of action(3) is also the gauge symmetry. 

To construct the massless fermionic 
modes of the Ramond sector, we equate the creation operator $d_{-1} = $
$\Gamma_k = (d_k + d_k^\dagger)$ for $k = 1, \cdots 5$; $\Gamma_k =
(d_{k-5}  - d_{k - 5}^{\dagger})/i$ for $k = 6, \cdots 10$ and for 
$k = 11$, $\Gamma_k = (d_{11} + d_{11}^{\dagger})$. These $\Gamma$'s 
obey the anticommutator relation
\begin{equation}
 \{\Gamma_k , \Gamma_l \} = 2 \delta_{k,l} 
\end{equation}
and also
\begin{equation}
 \Gamma_{11} = \Gamma_{1} \cdots \Gamma_{10}  
\end{equation}
Each $\Gamma$ is a $2^5 \times 2^5$ matrix and the $SO(11)$ spinors
are {\bf 32} dimensional. The massless spinors are
\begin{equation}
 \Psi_\alpha = \Gamma_{\alpha \beta}^{k} \mid 0 \rangle
{u}_{\beta}^{k}, \; \; \; \;  \alpha, \beta = 1, \cdots 32  
\end{equation}
The spinors decompose as follows:
when $SO(11) \rightarrow SO(6) \times SO(5)$; ${\bf 32} = (4, 4) + (\bar{4}, 4)$
and when $SO(11) \rightarrow SU(4) \times SU(2) \times SU(2), \; \;
{\bf 32} =
(4,2,1) + (4,2,1) + (\bar{4}, 2, 1) + (\bar{4},2,1)$. Finally in the
descent to the standard model
\begin{equation}
 SO(11) \rightarrow SU(3) \times SU_{L} (2) \times U_{Y} (1) 
\end{equation}
${\bf 32} = (3,2,1/3) + (3,1,4/3) + (3,1, -2/3) + (1,2,-1) + (1,1,-2) +
e_L$,  and $\nu_R$. $\nu_R$ is an addition to the usual standard
model. Thus we show that the fermionic spectrum are derivable as
string excitations directly in this model.
One of the main motivation of constructing this superstring is to show 
that the internal symmetry group makes a direct contact with the
standard model  which explains all available experimental
data with a high degree of accuracy. As already noted, the internal 
symmetry group was $SO(44)$. We divided these
fermions in groups of eleven where each group was characterised by a
space-time index $\mu = 0, 1, 2, 3$. All the four groups are similar,
but not idential.The group with $\mu = 0$ is different and physically 
absent. The other three groups of eleven, $\mu = 1, 2, 3$ are all identical.

$SO(6)$ and $SO(5)$ are the maximal subgroups of $SO(11)$
\cite{bd12}. Similarly $SU(3)$ and $U(1)$, two $SU(2)$s are the maximal
subgroups of $SO(6) \equiv SU (4)$ and $SO(5)$ respectively. So by the 
use of Wilson lines, the gauge group $SO(11)$ can break to the Pati
Salam group \cite{23}, $SU(4) \times SU_L (2) \times SU_R (2)$ and then
descent to $SU(3) \times U(1) \times SU(2) \times SU(2)$ without
breaking sypersymmetry. Thus the string is directly related to low
energy groups accessible to phenomology. This is an interesting
feature of the model. Departing from stringiness, let us record briefly some
consequences of phenomenology which are not found in literature. The most 
convinent scheme of descending to the standard model is
\begin{eqnarray}
SO(11) \longrightarrow && SO(6) \times SO(5) \nonumber \\
&& \downarrow M_X \nonumber \\
&& SU(4) \times SU_L (2) \times SU_R (2) \nonumber \\
&& \downarrow M_R \nonumber \\
&& \times \nonumber \\
&& \downarrow M_S \nonumber \\
&& SU_C (3) \times SU_L (2) \times U_Y (1)
\end{eqnarray}

Such a scheme and similar ones have been extensively studied
\cite{24}. Invoking charge quantisation \cite{24}, $SU(4)$ may be broken to
$U_{B-L} (1) \times SU_C (3)$ and subsequently $U_{B-L} (1)$ may
squeeze with $SU_R (2)$ to yield $U_Y (1)$. Unification mass is $M_X
= M_{GUT}$, the left-right symmetry breaks at $M_R$ and
supersymmetry is broken at $M_S = M_{SUSY}$. The renormalisation
equations for the evolution of the coupling constants are easily
written down \cite{25}.

We denote $\alpha_i = g_i^2 / 4  \pi$ where $g_i$ is the constant
related to the $i^{th}$ group, $\alpha_G = g_{U}^{2} / 4 \pi$ where
$g_U$ is the coupling constant at the GUT energy and $t_{XY} =
\frac{1}{2\pi} \log_e M_X / M_Y$. The lowest order evolution equations 
are
\begin{equation}
\alpha_{3}^{-1} (M_Z) = \alpha_{G}^{-1} + b_3 t_{SZ} + b_{3s} t_{RS} +
  b_{4s} t_{XR}, 
\end{equation}
\begin{equation}
\alpha_{2}^{-1} (M_Z) = \alpha_{G}^{-1} + b_2 t_{SZ} + b_{2s} t_{RS} +
  b_{2s} t_{XR},
\end{equation}
and
\begin{equation}
\alpha_{1}^{-1} (M_Z) = \alpha_{G}^{-1} + b_1 t_{SZ} + b_{1s} t_{RS} +
  ( \frac{2}{5} b_{4s} + \frac{3}{5} b_{2s} ) t_{XR}.
\end{equation}
$b_i$ and $b_{is}$ are the well known non-susy and susy coefficients
of the $\beta$-function respectively. The experimental values at $M_Z
= 91.18$ GeV are taken to be \cite{26}
\begin{equation}
\alpha_{1}^{-1} = 59.036, \alpha_{2}^{-1} = 29.656, \alpha_{3}^{-1} =
7.69
\end{equation}
To these, we add the expected string unification value
\begin{equation}
M_X = M_{GUT} = M_{string} = g_U (5 \times 10^{17}) GeV
\end{equation}

We have four unknown quantities to calculate from the four known
values.

Notice that the quantities, $b_1 - 3/5 b_2 = 6, b_{1s} - 3/5 b_{2s} = 6,
b_3 = -7, b_{3s} = -3$ and $b_{4s} = -6$ are independent of the
required number of Higgs doublets. So we rewrite the above three
equations as
\begin{equation}
\alpha_{1}^{-1} - 3/5 \alpha_{2}^{-1} - 2/5 \alpha_{3}^{-1} = 8.8
t_{SZ} + 7.2 t_{RS}
\end{equation}
\begin{equation}
\alpha_{1}^{-1} - 3/5 \alpha_{2}^{-1} = 2/5 \alpha_{G}^{-1} + 6 t_{SZ} 
+ 6 t_{RS} - 2.4 t_{XR}
\end{equation}
\begin{equation}
\alpha_{3}^{-1} = \alpha_{G}^{-1} - 7 t_{SZ} - 3 t_{RS} - 6 t_{XR}
\end{equation}
The solutions are $M_{SUSY} = 5 \times 10^9 $~GeV, $M_R = 5 \times
10^{14}$~GeV, $M_X = 2.87 \times 10^{17}$GeV and $g^2_U = 0.566$. With the
value of $M_R$ found here, the mass of the left-handed tau neutrino
\cite{28} is calculated to be about $\frac{1}{25}~eV$. following references
\cite{24} and \cite{27}. We have used $m_{top} (M_R) \cong 140 GeV$ in the 
formula for the neutrino mass $m_{\nu \tau}$,
\begin{equation}
m_{\nu \tau} = - \frac{m_{\rm top}^{2}}{ M_R}
\end{equation}

\section{Curved metric and Einstein equation}

We shall work in the orthonormal gauge where the zwebian is constant
and the gravitino field vanishes. The background gravitational field
will be denoted by $g_{\mu \nu} (x)$. The supersymetric action is
guessed to be \cite{29} 
\begin{eqnarray}
 S = && - \frac{1}{2} \int d^2 \sigma \left \{ \left [ \partial_\alpha X^\mu
\partial_\alpha X^\nu + i  \psi^{\mu, j} \rho^\alpha \left (
  \partial_\alpha \psi^{\nu}_j + \Gamma_{\lambda \sigma}^{\nu}
  \partial_{\lambda} X \psi^{\sigma}_{j} \right ) \right. \right.
\nonumber \\
 && \left.  + i  \phi^{\mu, k} \rho^{\alpha} \left ( \partial_{\alpha}
   \phi^{\nu}_{k} + \Gamma_{\lambda \sigma}^{\nu} \partial_{\alpha}
   X^\lambda \phi^{\sigma}_{k} \right ) \right ] g_{\mu \nu} (X)
\\ \nonumber 
 && \left. - \frac{1}{6} R_{\mu \nu \lambda \sigma} (X) \left ( 
\psi^{\mu, j} \psi^{\lambda}_{j} \psi^{\nu,j^\prime} \psi^{\sigma}_{j^\prime} +
\phi^{\mu,k} \phi^{\lambda}_{k} \phi^{\nu,k^\prime} \phi^{\sigma}_{k^\prime}
\right ) \right \} 
\end{eqnarray}
where the metric connection, the Christoffel symbol is
\begin{equation}
\Gamma_{\nu \lambda}^{\mu} = \frac{1}{2} g^{\mu \sigma} (
\partial_\lambda g_{\sigma \nu} + \partial_{\nu} g_{\sigma \lambda} -
\partial_{\sigma} g_{\lambda \nu} ) 
\end{equation}
and the Riemann Tensor is
\begin{equation}
R_{\nu \lambda \sigma}^{\mu} = \partial_{\sigma} \Gamma_{\nu
  \lambda}^\mu - \partial_\lambda \Gamma_{\nu\sigma}^{\mu} +
\Gamma_{\nu\lambda}^{m} \Gamma_{\sigma m}^{\mu} - \Gamma_{\nu
  \sigma}^{m} \Gamma_{\lambda m}^{\mu}
\end{equation}
Indeed, we find by lengthy and tedious calculation that the above
action is invariant under the followign global supersymmetric
transformations
\begin{equation}
\delta X^\mu = \bar{\epsilon} (  e^j_\psi \psi^{\mu}_{j} + i 
e^k_\phi \phi^{\mu}_{k}  )
\end{equation}
\begin{equation}
\delta \psi^{\mu,j} = - i e^j_\psi \rho^\alpha \partial_\alpha X^\mu \epsilon
- \Gamma_{\nu \lambda}^{\mu} (\bar{\epsilon} \psi^{\nu,j^\prime})
\psi^{\lambda}_{j^\prime} e_\psi^j
\end{equation}
\begin{equation}
\delta \phi^{\mu,k} = e_\phi^k \rho^\alpha \partial_\alpha X^\mu \epsilon - i
\Gamma_{\nu \lambda}^{\mu} (\bar{\epsilon}
\phi^{\nu,k^\prime})\phi^{\lambda}_{k^\prime} e_\phi^k
\end{equation}
We proceed to calculate the one-loop beta function or equivalently the 
one loop counter term by expanding the curved metric round $X^\mu =
X_{0}^{\mu}$. Replacing $X^\mu \rightarrow X^\mu_0 + X^\mu$, we have
the perturbative expansion
\begin{equation}
g_{\mu \nu} (X^\rho) = \eta_{\mu \nu} - \frac{1}{3} R_{\mu \nu \lambda 
  k} (X_0^\mu) X^{\lambda} X^k
\end{equation}
Eventually the lowest order counter term is obtained by contracting
two of the $X^\mu$'s that appear in the action equation. The quadratically
divergent integrals that arise due to contraction of terms like
$\langle \partial_\lambda X^\mu \partial_\lambda X^\nu \rangle$ are
discarded in dimensional regularisation. The logarithmic divergence in 
$2 + \epsilon$ dimension is
\begin{equation}
\langle X^\lambda (\sigma) X^k
(\sigma^\prime)\rangle_{\sigma\rightarrow \sigma^\prime} = \pi
\eta^{\lambda k} \int \frac{dk^{2+\epsilon}}{(2\pi)^{2+\epsilon}}
\frac{1}{k^2} \sim \frac{\eta^{\lambda k}}{2\epsilon}
\end{equation}
The infinite term in the one loop effective action is, thus,
\begin{equation}
\Delta S = - \frac{1}{12 \pi \epsilon} \int d^2 \sigma (
\partial_\lambda X^\mu \partial^\alpha X^\nu + i  \psi^{\mu,j}
\rho^{\alpha} \partial_\alpha \psi^{\nu}_{j} + i  \phi^{\mu,k}
\rho^{\alpha} \partial_\alpha \phi^{\nu}_{k} ) R_{\mu \nu} (X^\rho)
\end{equation}

$ R_{\mu \nu} (X^\rho)$ is the Ricci tensor. Vanishing of this term
which is needed for a finite theory or the vanishing of the one loop
$\beta$-function, namely
\[  R_{\mu \nu} (X^\rho) = 0 \]
are the familar Einstein equations of General Relativity in vacuum.

\section{Conclusion}
It is remarkable that we have been able to discuss physics from the Planck
scale to the Kamiokonda neutrino scale within the same framework. The
starting point has been a Nambu-Goto string in four dimensions to
which forty four Majorana fermions in four groups have been
added. The resulting string has an action which is world sheet 
supersymmetric. Super-Virasoro algebra for the energymomentum tensor and 
current generators is established. Conformal ghosts are introduced
whose contributions cancel the anomalies. BRST charge is explicitly 
constructed. Since the expectation value of the Hamiltonian is deduced
to be the sum of squares of the supersymmetric charges, the tachyons
should be absent. Thus we have been
able to  show that the $26 D$ ordinary string 
behaves like  a $4 D$ superstring. This paper is to elaborate the material 
contained in~\cite{30}.

By the use of Wilson lines, the gauge symmetry $SO(11)$ can break to
Pati Salam group and $SU_C (3) \times SU_L (2) \times SU_R (2) \times
U (1)$ preserving supersymmetry. The left right
symmetry and supersymmetry are broken at intermediate mass scales. 
By the usual see-saw mechanism, the left handed neutrino develops a
small mass of  about $\frac{1}{25}$ ev. Finally
the descent is complete at $SU_C (3) \times SU_L (2) \times U_Y
(1)$. There is no gap left between $M_{GUT}$ and $M_{string}$ by
choice.\\ 
A global supersymmetric action in orthonormal gauge has been 
constructed in a gravitational background field. The vanishing of the
one loop divergent constribution to 
the action require that the Ricci tensor vanishes which are the 
Einsteins equations of general relativity in vacuum. 

We have profited from discussions with Dr. J.  Maharana and Dr. S. Mahapatra.
We thank Sri D. Pradhan for computer compilation and the Institute of
Physics for providing Library and  Computer  facilities.

\today\hfill boss.tex

\begin{thebibliography}{9999}
\bibitem{bd1} Y. Nambu, in {\it Symmetries and quark models},
ed. R. Chand (Gordon and Breach), (1970);
H. B. Niclson, $15^{th}$ International Conference on
High Energy Physics, Kiev (1970);
L. Susskind, Phys. Rev. {\bf D1}, 1182  (1970).

\bibitem{bd2} R. Dolen, D. Horn and C. Schmid, Phys. Rev. Lett.
{\bf 19}, 402 (1967) ; Phys. Rev. {\bf 166},  1768 (1968);
G. Veneziano, Nuovo. Cim {\bf 57A},  190 (1968).

\bibitem{bd3} C. Lovelace, Phys. Lett. {\bf 34B}, 500  (1971).

\bibitem{gates} S. James Gates, Jr. and W. Siegel, Phys. Lett.
{\bf B20},631(1988); D. A. Depireux, S. James Gates, Jr. and Q-Han
Park, 
Phys. lett.{\bf B224}, 364(1989); S. Bellucci, D. A. Depireux and 
S. James Gates, Jr. Phys. Lett.{\bf B232}, 67(1989);
D. A. Depireux, S. James Gates, Jr. and B. Radak, Phys. Lett.{\bf B236}, 
411(1990).
\bibitem{bd99} {\it String theory in four dimensions} edited by 
M. Dine, North Holland, (1988).

\bibitem{ia} I. Antoniadis, C. Buchas and C. Kounnas, 
Nucl. Phys. {\bf B289},  87 (1987), D. Chang and A. Kumar,
Phys. Rev. {\bf D38}, 1893  (1988); ibid , {\bf D38}, 3739  (1988),  
H. Kawai, D. C. Lewellen and C-H. H. Tye, Nucl. Phys. {\bf B288},  1 (1987).
\bibitem{6} A. Casher, F. Englert, H. Nicolai and A. Taormina,
  Phys. LEtt. {\bf B 162}, 121 (1985).
\bibitem{7} F. Englert, L. Houart and A. Taormina, JHEP, 0108:013, 1 (2001).

\bibitem{bd11} T. H. R. Skyrme, Proc. Roy. Soc. 
{\bf A262}, 237  (1961).

\bibitem{bd5}  M. B. Green. J. H. Schwarz and E. Witten,
{\it Superstring Theory} Vol. 1, Cambridge University Press, 
Cambridge, England (1987), references to the materials 
presented in this work can be found in this book.

\bibitem{bd4}A. Neveu and J. H. Schwarz, Nucl Phys.
{\bf B31}, 86 (1971)  ; A. Neveu, J. H. Schwarz and C. B.
Thorn, {\it ibid} {\bf 35B},  529  (1971).

\bibitem{bd6} M. A. Virasoro, Phys. Rev. {\bf D1},  2933  (1970).

\bibitem{bd7} L. D. Faddeev and V. N. Popov, Phys. Lett
{\bf 25B},29  (1967)  ; A. M. Polyakov, Phys. Lett. {\bf 103B}, 
 (1981), {\it ibid} {\bf 103B},  211 (1981).
\bibitem{bd8} C. Becchi, A. Rouet and R. Stora, Phys. Lett.
{\bf 52B},  344 (1974); Ann. Phys. {\bf 98}, 287  (1976);
I. V. Tyupin, Lebedev Preprint FIAN No. {\bf 39}, (1975). 
\bibitem{ramond} P. Ramond, Phys. Rev. {\bf D3}, 2415  (1971).
\bibitem{bd9} D. Friedman, E. Martinee and S. Shanker, Nucl.
Phys. {\bf B271}, 93  (1986).
\bibitem{bd10} F. Gliozzi, J. Scherk and D. Olive, Phys. Lett.
{\bf 65B}, (1976) 282   and references there in ;
Nucl. Phys. {\bf B122},253 (1977). 
\bibitem{bd13} M. Kaku in {\it Strings, Conformal 
Fields and Topology }, Sringer-Verlag, New York(1991) and Introduction 
to Superstrings, Springer-Verlag, New York (1988).
\bibitem{bd98} N. Seiberg and E. Witten, Nucl. Phys. {\bf B276}, 27  (1986).
\bibitem{bd97} A. Erdelyi et al., {\it Higher Transcendetal Functions},
Mc Graw Hill, New York, (1953).
\bibitem{bd96}  see for instance, the introduction given in
{\it Superstring Construction}, edited by B. Schellekens (North Holland, 1989).  
\bibitem{bd12} R. Slansky, Phys. Rep. {\bf 79},1 (1981) .
\bibitem{bd22} B. B. Deo, Mod. Phys. Lett. {\bf A 13}, 2971 (1998).
\bibitem{23} J. C. Pati and A. Salam, Phys. Rev. {\bf D10}, 275
(1979), R. N. Mohapatra and J. C. Pati, Phys. Rev. {\bf D11}, 2558  (1975).
\bibitem{24} R. N. Mohapatra, {\it Unification and Supersymmetry},
Springer-Verlag, New York (1986); P. Langaker and S. Uma Sankar, 
Phys. Rev. {\bf D40},1385  (1989) .
\bibitem{25} F. D. Aguila and L. E. Ibanez, Nucl. Phys. {\bf B177}, 
60  (1981), C. S. Aulakh 
and R. N. Mohapatra Phys. Rev. {\bf D28}, 217   (1983).
\bibitem{26} P. Langacker and N. Polonsky, Phys. Rev. {\bf D 52}, 3081  (1995).
\bibitem{27} J. C. Pati, Intl. Jr. Mod. Phys. {\bf A 14}, 2949  (1999).
\bibitem{28} Super-Kamiokanda Collaboration, Y. Fakuda et. al., 
Phys. Rev. Lett. {\bf 81},1562 (1998).\bibitem{29} A. Das, J. Maharana 
and S. Roy, Phys. Rev. {\bf D40},2037  (1989), S. Fubini, 
J. Maharana, M. Roncadelli and G. Veneziano, Nucl. Phys. {\bf B316},36  (1989).
\bibitem{30} B. B. Deo, Phys. Lett. {\bf B 557}, 115 (2003),\\
B. B. Deo and L. Maharana, ` {\it Derivation of Einstein Equation from a 
new type of superstring in four dimension}' hep-th/0212004,\\ 
B. B. Deo, `{\it Suppersymmetric 
Standard model from String theory}, hep-th/0301017,\\ B. B. Deo and L. Maharana,
{\it Particle Spectrum of the Supersymmetric Standard Model from the 
Massless Excitations of a Four Dimensional Superstring}, hep-th/0302133
\end{thebibliography}
\end{document}